\documentclass[runningheads]{llncs}
\usepackage{graphicx}
\usepackage{graphicx}
\usepackage{color}
\usepackage{longtable}
\usepackage{booktabs}
\usepackage{bm}
\usepackage{amsmath}
\usepackage[ruled,vlined]{algorithm2e}
\usepackage{lipsum}
\usepackage{mathtools}
\usepackage{cuted}
\usepackage{flushend}
\usepackage{gensymb}
\usepackage{float}
\usepackage{enumitem}
\usepackage{amsfonts}

\begin{document}

\title{Lymph Node Gross Tumor Volume Detection and Segmentation via Distance-based Gating using 3D CT/PET Imaging in Radiotherapy}
\author{Zhuotun Zhu\inst{1,2}\and Dakai Jin\inst{1}\and Ke Yan\inst{1}\and Tsung-Ying Ho\inst{3}\and Xianghua Ye\inst{5}\and Dazhou Guo\inst{1}\and Chun-Hung Chao\inst{4}\and Jing Xiao\inst{6}\and Alan Yuille\inst{2}\and Le Lu\inst{1}}
%index{Zhu, Zhuotun}
%index{Jin, Dakai}
%index{Yan, Ke}
%index{Ho, Tsung-Ying}
%index{Ye, Xianghua}
%index{Guo, Dazhou}
%index{Chao, Chun-Hung}
%index{Xiao, Jing}
%index{Yuille, Alan}
%index{Lu, Le}

\authorrunning{Z. Zhu et al.}

\institute{
PAII Inc., Bethesda MD, USA \and Johns Hopkins University, Baltimore MD, USA
\and
Chang Gung Memorial Hospital, Linkou, Taiwan, ROC \and
National Tsing Hua University, Hsinchu City, Taiwan, ROC\and
The First Affiliated Hospital Zhejiang University, Hangzhou, China
\and 
Ping An Technology, Shenzhen, China}

%\institute{Anonymous Institute}
\titlerunning{GTVLN Detection and Segmentation via Distance-based Gating}
\maketitle              % typeset the header of the contribution
\begin{abstract}
Finding, identifying and segmenting suspicious cancer metastasized lymph nodes from 3D multi-modality imaging is a clinical task of paramount importance. In radiotherapy, they are referred to as Lymph Node Gross Tumor Volume (GTV$_{LN}$). Determining and delineating the spread of GTV$_{LN}$ is essential in defining the corresponding resection and irradiating regions for the downstream workflows of surgical resection and radiotherapy of various cancers. In this work, we propose an effective distance-based gating approach to simulate and simplify the high-level reasoning protocols conducted by radiation oncologists, in a divide-and-conquer manner. GTV$_{LN}$ is divided into two subgroups of ``tumor-proximal" and ``tumor-distal", respectively, by means of binary or soft distance gating. This is motivated by the observation that each category can have distinct though overlapping distributions of appearance, size and other LN characteristics. A novel multi-branch detection-by-segmentation network is trained with each branch specializing on learning one GTV$_{LN}$ category features, and outputs from multi-branch are fused in inference. The proposed method is evaluated on an in-house dataset of $141$ esophageal cancer patients with both PET and CT imaging modalities. Our results validate significant improvements on the mean recall from $72.5\%$ to $78.2\%$, as compared to previous state-of-the-art work. The highest achieved GTV$_{LN}$ recall of $82.5\%$ at $20\%$ precision is clinically relevant and valuable since human observers tend to have low sensitivity ($\sim80\%$ for the most experienced radiation oncologists, as reported by literature \cite{goel2017clinical}).

\end{abstract}

\section{Introduction}

Assessing the lymph node (LN) status in oncology clinical workflows is an indispensable step for the precision cancer diagnosis and treatment planning, e.g., radiation therapy or surgical resection. The class of enlarged LN is defined by the revised RECIST guideline \cite{schwartz2009evaluation} if its short axial axis is more than $10{\textrm{-}}15$ mm in computed tomography (CT). In radiotherapy treatment, both the primary tumor and all metastasis suspicious LNs must be sufficiently treated within the clinical target volume with the proper doses~\cite{jin2019deep}. We refer these LNs as lymph node gross tumor volume or GTV$_{LN}$, which includes enlarged LNs, as well as smaller ones that are associated with a high positron emission tomography (PET) signal or any metastasis signs in CT~\cite{scatarige1983low}. Accurately identifying and  delineating GTV$_{LN}$, to be spatially included in the treatment area, is essential for a desirable cancer treatment outcome~\cite{National2020}.

\begin{figure}[t]
\centering
\includegraphics[width=0.9\textwidth]{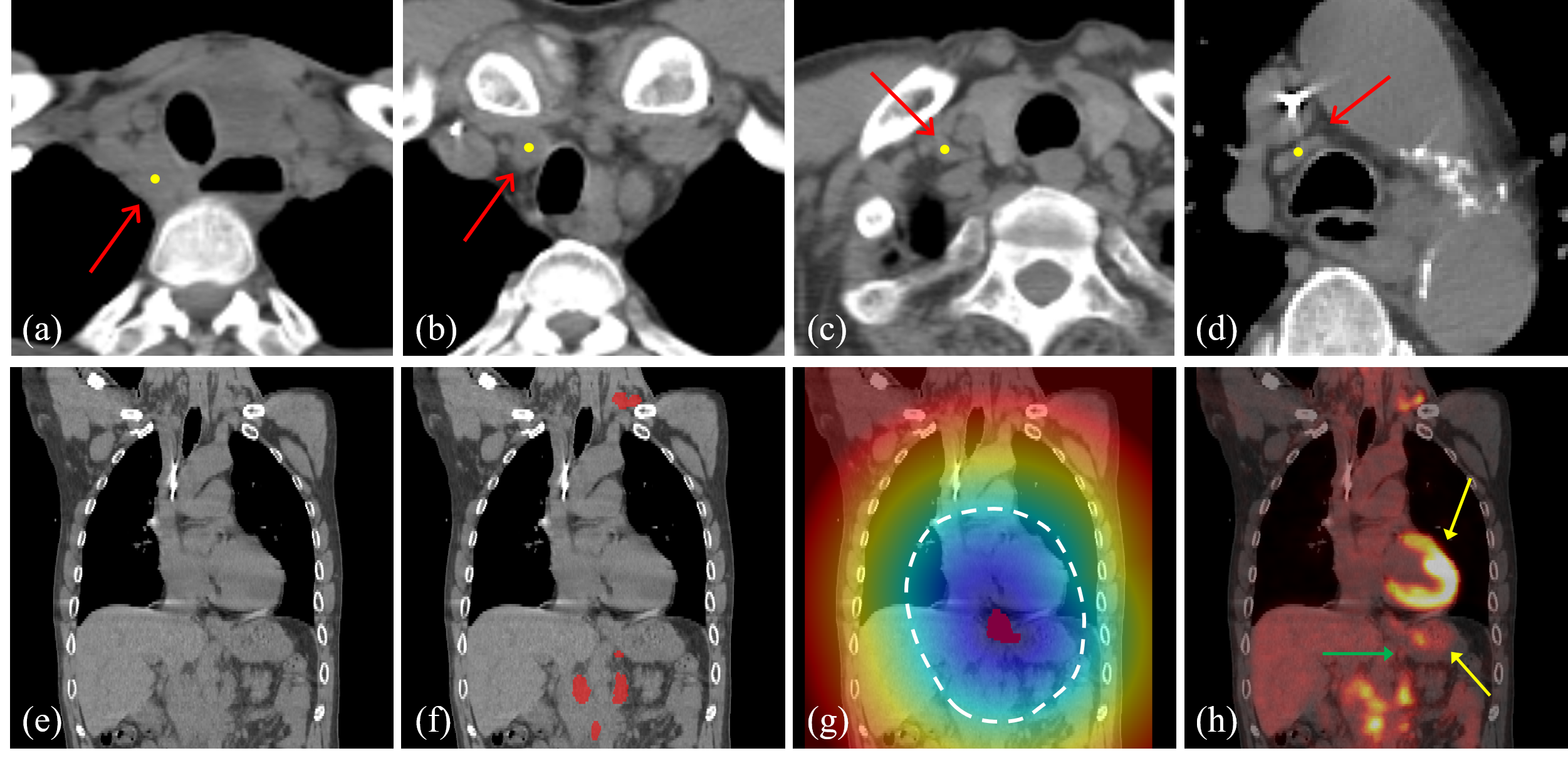}
\caption{Top row (a-d): examples of the GTV$_{LN}$ (red arrow) with varying size and appearance at scatteredly distributed locations. Bottom row (e-h): (e) A coronal view of RTCT for an esophageal cancer patient. (f) The manual annotated GTV$_{LN}$ mask. (g) The tumor distance transformation map overlaid on RTCT, where the primary tumor is indicated by red in the center and the white dash line shows an example of the binary tumor proximal and distal region division. (h) PET imaging shows several FPs with high signals (yellow arrows). Two FN GTV$_{LN}$ are indicated by green arrow where PET has even no signals on a GTV$_{LN}$.}\label{Fig:LNSamples}
\end{figure}

It is an extremely challenging and time-consuming task to identify GTV$_{LN}$, even for experienced radiation oncologists. High-level sophisticated clinical reasoning guidelines are needed, leading to the risk of uncertainty and subjectivity with high inter-observer variabilities~\cite{goel2017clinical}. It is arguably more difficult than detecting the more general enlarged LNs. (1) Finding GTV$_{LN}$ is often performed using radiotherapy CT (RTCT) that (unlike diagnostic CT) is not contrast-enhanced. Hence the metastasis signs for identifying GTV$_{LN}$ are subtler. (2) GTV$_{LN}$ itself has poor contrast. Because of the shape and appearance ambiguity, it can be easily confused with vessels or muscles. (3) The size and shape of GTV$_{LN}$ vary considerably with large amounts of smaller ones that are harder to detect.  Refer Fig.~\ref{Fig:LNSamples} (top row) for an illustration of GTV$_{LN}$. While many previous works attempt to detect enlarged LNs using contrast-enhanced CT  \cite{barbu2011automatic,bouget2019semantic,feulner2013lymph,nogues2016automatic,roth2015improving,roth2014new,yan2018deeplesion}, no work, as of yet, has studied the GTV$_{LN}$ detection in non-contrast RTCT scans. Given the evident differences between the enlarged LNs and GTV$_{LN}$, further innovations are required for the robust GTV$_{LN}$ detection and segmentation.

Valuable insights from physicians' clinical diagnosis and analysis process can be leveraged to tackle this problem. As one of the primary cues, human observers condition the analysis of GTV$_{LN}$ based on the LNs' distance with respect to the corresponding primary tumor location. For LNs proximal to the tumor, physicians more readily identify them as GTV$_{LN}$ in radiotherapy treatment. However, for LNs distal to the tumor, they use more strict criteria to include if there are clear signs of metastasis,  {\it{e.g.}}, enlarged size, increased PET signals, and/or other CT based evidence~\cite{scatarige1983low}. Hence, the distance measure relative to the primary tumor plays a key role during physician's decision making. Besides the distance, the PET modality is also of high importance. Although as a noisy imaging channel, it has shown to be helpful in increasing the GTV$_{LN}$ detection sensitivity~\cite{goel2017clinical}. As demonstrated in Fig.~\ref{Fig:LNSamples} (bottom row), PET provides critically distinct information, yet, it also exhibits false positives (FPs) and false negatives (FNs).

In this paper, we imitate the physician's diagnosis process to tackle the problem of GTV$_{LN}$ detection and segmentation. (1) We introduce a distance-based gating strategy in a multi-task framework to divide the underlying GTV$_{LN}$ distributions into ``tumor-proximal" and ``tumor-distal" categories and solve them accordingly. Specifically, a multi-branch network is proposed to adopt a shared encoder and two separate decoders to detect and segment the ``tumor-proximal" and ``tumor-distal" GTV$_{LN}$, respectively. A distance-based gating function is designed to generate the corresponding GTV$_{LN}$ sample weights for each branch. By applying the gating function at the outputs of decoders, each branch is specialized to learn the ``tumor-proximal" or ``tumor-distal" GTV$_{LN}$ features that emulates physician's diagnosis process. (2) We leverage the early fusion (EF) of three modalities as input to our model, {\it{i.e.}}, RTCT, PET and 3D tumor distance map (Fig.~\ref{Fig:LNSamples}(bottom row)). RTCT depicts anatomical structures capturing the intensity, appearance and contextual information, while PET provides metastasis functional activities. Meanwhile, the tumor distance map further encodes the critical distance information in the network. Fusion of these three modalities together can effectively boost the GTV$_{LN}$ identification performance. (3) We evaluate on a dataset comprising $651$ voxel-wise labeled GTV$_{LN}$ instances in $141$ esophageal cancer patients, as the largest GTV$_{LN}$ dataset to date for chest and abdominal radiotherapy. Our method significantly improves the detection mean recall from $72.5\%$ to $78.2\%$, compared with the previous state-of-the-art lesion detection method \cite{yan2019mulan}. The highest achieved recall of $82.5\%$ is also clinically relevant and valuable. As reported in \cite{goel2017clinical}, human observers tend to have relatively low GTV$_{LN}$ sensitivities, {\it{e.g.}},  $\sim80\%$ by even very experienced radiation oncologists. This demonstrates our work's clinical values. 

\section{Method}

\begin{figure}[t]
\centering
\includegraphics[width=1.0\textwidth]{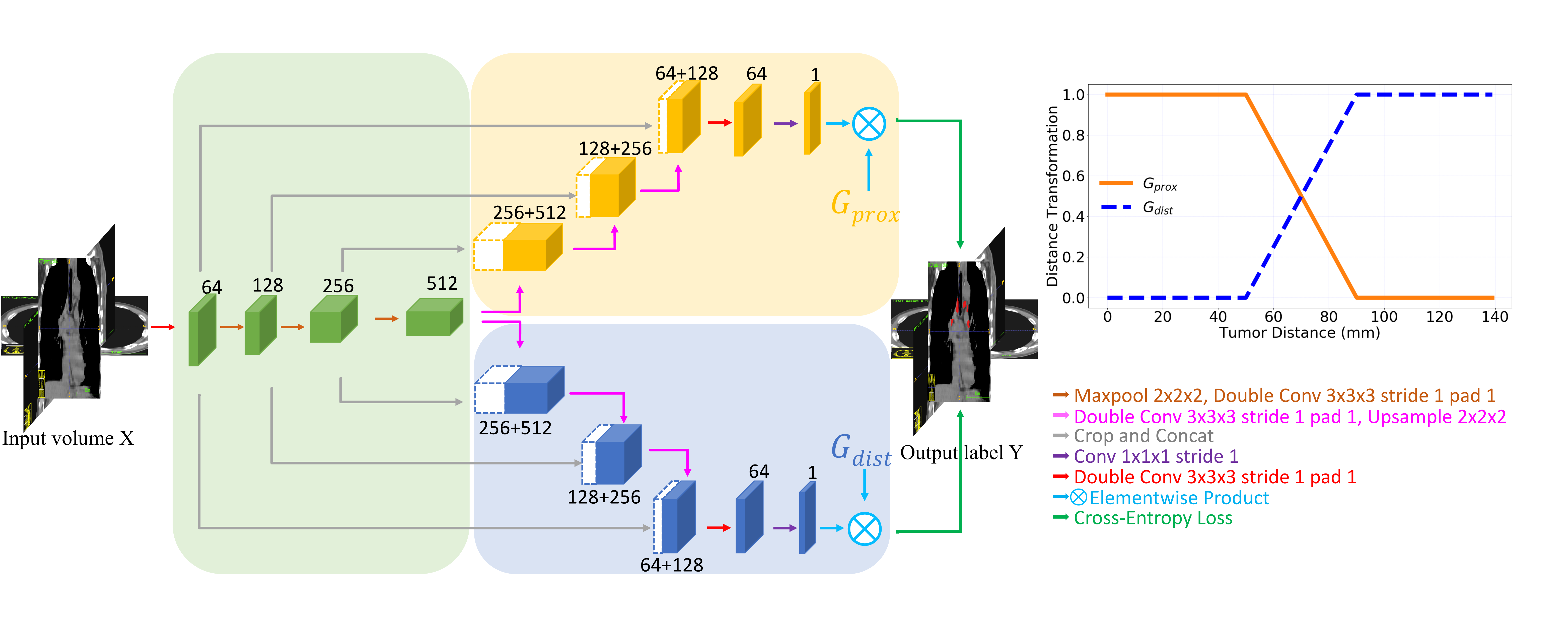}
\caption{The overall framework of our proposed multi-branch GTV$_{LN}$ detection and segmentation method. The light green part shows the encoder path, while the light yellow and light blue parts show the two decoders, respectively. The number of channels is denoted either on the top or the bottom of the box.} \label{Fig:Flowchart}
\end{figure}

Fig.~\ref{Fig:Flowchart} shows the framework of our proposed multi-branch GTV$_{LN}$ detection-by-segmentation method. Similar to~\cite{zhu2020segmentation,zhu2019multi} which are designed for the pancreatic tumors, we detect GTV$_{LN}$ by segmenting them. We first compute the 3D tumor distance transformation map (Sec. \ref{sec:Distmap}), based on which any GTV$_{LN}$ is divided into the tumor-proximal or tumor-adjacent subcategory. Next, a multi-branch detection-by-segmentation network is designed where each branch focuses on one subgroup of GTV$_{LN}$ segmentation (Sec. \ref{sec:Multi}). This is achieved by applying a binary or soft distance-gating function imposed on the penalty function at the output of the two branches (Sec. \ref{Sec:Gating}). Hence, each branch can learn specific parameters to specialize on segmenting and detecting the tumor-proximal and tumor-adjacent GTV$_{LN}$, respectively.

\subsection{3D Tumor Distance Transformation}\label{sec:Distmap}
To stratify GTV$_{LN}$ into tumor-proximal and tumor-distal subgroups, we first compute the 3D tumor distance transformation map, denoted as $\mathbf{X}^{\mathrm{D}}$, from the primary tumor $\mathcal{O}$. The value at each voxel $x_i$ represents the shortest distance between this voxel and the mask of the primary tumor. Let $B(\mathcal{O})$ be a set that includes the boundary voxels of the tumor. The distance transformation value at a voxel $x_i$ is computed as

\begin{align}
\mathbf{X}^{\mathrm{D}}(x_i) = \left \{
\begin{array}{rcl}
\underset{q\in B(\mathcal{O})}{\min} d(x_i,q)  & \quad {\text{if} \quad x_i\notin \mathcal{O}}\\
0  & \quad {\text{if} \quad x_i\in \mathcal{O} }
\end{array} \right. \mathrm{,}
\end{align}
where $d(x_i,q)$ is the Euclidean distance from $x_i$ to $q$. $\mathbf{X}^{\mathrm{D}}$ can be efficiently computed using algorithms such as the one proposed in~\cite{maurer2003linear}. Based on $\mathbf{X}^{\mathrm{D}}$, GTV$_{LN}$ can be divided into tumor-proximal and tumor-distal subgroups using either binary or soft distance-gating function as explained in detail in Sec.~\ref{Sec:Gating}.

\subsection{Multi-branch Detection-by-Segmentation via Distance Gating} \label{sec:Multi}
GTV$_{LN}$ identification is implicitly associated with their distance distributions to the primary tumor in the diagnosis process of physicians. Hence, we divide GTV$_{LN}$ into tumor-proximal and tumor-distal subgroups and conduct detection accordingly. To do this, we design a multi-branch detection-by-segmentation network with each branch focusing on segmenting one GTV$_{LN}$ subgroup. Each branch is implemented by an independent decoder to learn and extract the subgroup specific information, while they share a single encoder to extract the common GTV$_{LN}$ image features. Assuming there are $N$ data samples, we denote a dataset as $\mathbf{S}=\left\{\left(\mathbf{X}^{\mathrm{CT}}_n, \mathbf{X}^{\mathrm{PET}}_n, \mathbf{X}^{\mathrm{D}}_n,  \mathbf{Y}_n\right)\right\}_{n=1}^{N}$, where $\mathbf{X}^{\mathrm{CT}}_n$, $\mathbf{X}^{\mathrm{PET}}_n$, $\mathbf{X}^{\mathrm{D}}_n$ and $\mathbf{Y}_n$ represent the non-contrast RTCT, registered PET, tumor distance transformation map, and ground truth GTV$_{LN}$ segmentation mask, respectively. Without the loss of generality, we drop $n$ for conciseness in the rest of this paper. The total number of branches is denoted as $M$, where $M=2$ in our case. A CNN segmentation model is denoted as a mapping function $\mathbb{E}: \mathbf{P} = {\mathbf{f}\left(\mathbf{\mathcal{X}}; \boldsymbol{\Theta}\right)}$, where $\mathbf{\mathcal{X}}$ is a set of inputs, which consists of a single modality or a concatenation of multiple modalities. $\boldsymbol{\Theta}$ indicates model parameters, and $\mathbf{P}$ means the predicted probability volume.  Given that $p(y_i | x_i; \boldsymbol{\Theta}_m)$ represents the predicted probability of a voxel $x_i \in\mathbf{\mathcal{X}}$ being the labeled class from the $m$th branch, the overall negative log-likelihood loss aggregated across $M$ branches can be formulated as:

\begin{equation}\label{Eq:MultiSegLoss}
\mathcal{L} = \sum_m{\mathcal{L}_m(\mathbf{\mathcal{X}}}; \boldsymbol{\Theta}_m, \boldsymbol{G}_m) = -\sum_{i}\sum_m{{g_{m,i}\log(p(y_i | x_i; \boldsymbol{\Theta}_m))}},
\end{equation}
where $\boldsymbol{G}={\left\{\boldsymbol{G}_m\right\}}_{m=1}^{M}$ is introduced as a set of volumes containing the transformed gating weights at each voxel based on its distance to the primary tumor. At every voxel $x_i \in \boldsymbol{G}$, the gating weights satisfies $\sum_m{g_{m,i}} = 1$. 

\subsection{Distance-based Gating Module}\label{Sec:Gating}
Based on the tumor distance map $\mathbf{X}^{\mathrm{D}}$, our gating functions can be designed to generate appropriate GTV$_{LN}$ sample weights for different branches so that each branch specializes on learning the subgroup specific features. In our case, we explore two options: (1) binary distance gating and (2) soft distance gating. 

{\bf Binary Distance Gating (BG).} Based on the tumor distance map $\mathbf{X}^{\mathrm{D}}$, we divide image voxels into two groups, $x_\mathrm{prox}$ and $x_\mathrm{dis}$, to be tumor-proximal and tumor-distal, respectively, where $\mathrm{prox}=\{i|x_i^{\mathrm{D}}\leq d_0, x_i^{\mathrm{D}}\in\mathbf{X}^{\mathrm{D}} \}$ and $\mathrm{dis}=\{i|x_i^{\mathrm{D}}>d_0, x_i^{\mathrm{D}}\in\mathbf{X}^{\mathrm{D}}\}$. Therefore the gating transformations for two decoders are defined as $\mathbf{\boldsymbol{G}_{prox}}=\mathbf{1}[x_i^{\mathrm{D}}\leq d_0]$ and $\mathbf{\boldsymbol{G}_{dist}}=1-\mathbf{\boldsymbol{G}_{prox}}$, where $\mathbf{1}[\cdot]$ is an indicator function which equals one if its argument is true and zero otherwise. In this way, we divide the GTV$_{LN}$ strictly into two disjoint categories, and each branch focuses on decoding and learning from one category. 

{\bf Soft Distance Gating (SG).} We further explore a soft gating method that linearly changes the penalty weights of GTV$_{LN}$ samples as they are closer or further to the tumor. This can avoid a sudden change of weight values when samples are near the proximal and distal category boundaries. Recommended by our physician, we formulate a soft gating module based on $\mathbf{X}^{\mathrm{D}}$ as following:
\begin{align}\label{Eq:SG}
\mathbf{G_{prox}}(x_i) = \left \{
\begin{array}{rll}
1 - \frac{x_i^{\mathrm{D}}-d_{prox}}{d_{dist} - d_{prox}}  & \quad {\text{if} \quad d_{prox}< x_i^{\mathrm{D}}\leq d_{dist} }\\
1  & \quad {\text{if} \quad x_i^{\mathrm{D}}\leq d_{prox}}\\
0  & \quad {\text{if} \quad x_i^{\mathrm{D}}> d_{dist}}
\end{array} \right. \mathrm{,}
\end{align}
and $\mathbf{G_{dist}}(x_i)=1-\mathbf{G_{prox}}(x_i)$ accordingly. 

\section{Experimental Results}
\subsection{Dataset and Preprocessing}
{\bf Dataset.} We collected $141$ non-contrast RTCTs of esophageal cancer patients, with all undergoing radiotherapy treatments. Radiation oncologists labeled 3D segmentation masks of the primary tumor and all GTV$_{LN}$. For each patient, we have a non-contrast RTCT and a pair of PET/CT scans. There is a total of $651$ GTV$_{LN}$ with voxel-wise annotations in the mediastinum or upper abdomen regions, as the largest annotated GTV$_{LN}$ dataset to-date. We randomly split patients into $60\%$, $10\%$, $30\%$ for training, validation and testing, respectively. 

{\bf Implementation Details.}
In our experiments, PET scan is registered to RTCT using the similar method described in~\cite{jin2019accurate}. Then all coupling pairs of RTCT and registered PET images are resampled to have a consistent spatial resolution of $1\times1\times2.5$ mm. To generate the 3D training samples, we crop sub-volumes of $96\times96\times64$ from the RTCT, registered PET and the tumor distance map around each GTV$_{LN}$ as well as randomly from the background. For the distance-gating related parameters, we set $d_0=7$ cm as the binary gating threshold, and $d_{prox}=5$ cm and $d_{dist}=9$ cm as the soft gating thresholds, respectively, as suggested by our clinical collaborator. We further apply random rotations in the x-y plane within $10$ degrees to augment the training data.

Detection-by-segmentation models are trained on two NVIDIA Quadra RTX 6000 GPUs with a batch size of $8$ for $50$ epochs. The RAdam~\cite{liu2019variance} optimizer with a learning rate of $0.0001$ is used with a momentum of $0.9$ and a weight decay of $0.0005$. For inference, 3D sliding windows with a sub-volume of $96\times96\times64$ and a stride of $64\times64\times32$ voxels are processed. For each sub-volume, predictions from two decoders are weighted and aggregated according to the gating transformation $\mathbf{G_m}$ to obtain the final GTV$_{LN}$ segmentation results.

{\bf Evaluation Metrics.} 
We first describe the hit criteria, {\it{i.e.}}, the correct detection, for our detection-by-segmentation method. For an GTV$_{LN}$ prediction, if it overlaps with any ground-truth GTV$_{LN}$, we treat it as a hit provided that its estimated radius is similar to the radius of the ground-truth GTV$_{LN}$ within the range of $[0.5, 1.5]$. The performance is assessed using the mean and max recall (mRecall and Recall$_{max}$) at a precision range of $[0.10, 0.50]$ with $0.05$ interval, and the mean free response operating characteristic (FROC) at $3,4,6,8$ FPs per patient. These operating points were chosen after confirming with our physician. 

{\bf Comparison Setups.}
Using the binary and soft distance-based gating function, our multi-branch GTV$_{LN}$ detection-by-segmentation method is denoted as {\bf multi-branch BG} and {\bf multi-branch SG}, respectively. We compare against the following setups: (1) a single 3D UNet~\cite{cciccek20163d} trained using RTCT alone or the early fusion (EF) of multi-modalities (denoted as {\bf single-net} method); (2) Two separate UNets trained with the corresponding tumor-proximal and tumor-distal GTV$_{LN}$ samples and results spatially fused together (our preliminary work~\cite{zhu2020detecting} denoted as {\bf multi-net BG}); and (3) MULAN~\cite{yan2019mulan}, a state-of-the-art (SOTA) general lesion detection method on DeepLesion~\cite{yan2018deeplesion} that contains more than $10\textrm{,}000$ enlarged LNs. 

\subsection{Quantitative Results \& Discussion}\label{Sec:Eva}

Our quantitative results and comparisons are given in Table.~\ref{Tab:FullResults}. Several observations can be drawn on addressing the effectiveness of our proposed methods. {\bf (1)} The multi-modality input, {\it{i.e.}}, early fusion (EF) of RTCT, PET and tumor distance map, are of great benefits for detecting the GTV$_{LN}$. There are drastic performance improvements of absolute $6.7\%$ and $7.2\%$ in mRecall and mFROC when EF is adopted as compared to using RTCT alone. These results validate that input channels of PET functional imaging and 3D tumor distance transform map are valuable for identifying GTV$_{LN}$. {\bf (2)} The distance-based gating strategies are evidently effective as the options of {\bf multi-net BG}, {\bf multi-branch BG} and {\bf multi-branch SG} consistently increase the performance. For example, the multi-net BG model achieves $74.7\%$ mRecall and $69.5\%$ mFROC, which is a $1.6\%$ and $1.9\%$ improvement against the best single-net model (where no distance-based stratification is used). The performance further boosts with the network models of multi-branch BG and multi-branch SG, to the highest scores of $78.2\%$ mRecall and $72.4\%$ mFROC achieved by the multi-branch SG. 

\begin{table}[t]
\caption{Quantitative results of our proposed methods with the comparison to other setups and the previous state-of-the-art.}
\label{results}
\footnotesize
\centering
\begin{tabular}{l|cc|cc|ccc}
\hline
Methods:            &  CT    &  EF        &  mRecall  & Recall$_{max}$   & mFROC  & FROC@4  &FROC@6          \\ \hline
single-net          &    \checkmark    &    &  $0.664$ & $0.762$ & $0.604$    & $0.552$      &  $0.675$   \\ 
single-net      &   &   \checkmark    &  $0.731$ & $0.820$ & $0.676$     & $0.667$  & $0.713$  \\ \hline
multi-net BG~\cite{zhu2020detecting}    &   &   \checkmark  & $0.747$  & $0.825$  & $0.695$  &  $0.668$   &  $\bm{0.739}$  \\
multi-branch BG (Ours)    &   &   \checkmark   & $0.761$  & $\bm{0.845}$  & $0.679$    & $0.667$    &  $0.716$  \\
multi-branch SG (Ours)   &   &   \checkmark & $\bm{0.782}$  & $0.843$ & $\bm{0.724}$    & $\bm{0.729}$    & $0.738$  \\  \hline
MULAN~\cite{yan2019mulan}  &  \checkmark  &         & $0.711$  & $0.758$  & $0.632$    & $0.632$  &  $0.642$   \\
MULAN~\cite{yan2019mulan}  &   &   \checkmark      & $0.725$  & $0.781$  & $0.708$   & $0.718$   &  $0.720$ \\ \hline
\end{tabular}
\label{Tab:FullResults}
\end{table}

{\bf Multi-branch versus Multi-net.} Using the distance-based gating strategy, our proposed multi-branch methods perform considerably better than the {\bf multi-net BG} model. Even our second best model {\bf multi-branch BG}, the mean and maximal recalls have been improved by $1.4\%$ (from $74.7\%$ to $76.1\%$) and $2.0\%$ (from $82.5\%$ to $84.5\%$) against the {\bf multi-net BG} model. When the multi-branch framework is equipped with the {\bf soft-gating}, marked improvements of absolute $3.5\%$ and $2.9\%$ in both mRecall and mFROC are observed as compared against to the {\bf multi-net BG} model. This validates the effectiveness of our jointly trained multi-branch framework design, and our intuition that gradually changing GTV$_{LN}$ weights for the proximal and distal branches are more natural and effective. As we recall, the multi-net baseline directly trains two separate 3D UNets~\cite{cciccek20163d} targeted to segment each GTV$_{LN}$ subgroup. Considering the limited GTV$_{LN}$ training data (a few hundreds of patients), it can be overfitting prone from the split to even smaller patient subgroups. 

Table.~\ref{Tab:FullResults} also compares with the SOTA universal lesion detection method, i.e., MULAN~\cite{yan2019mulan} on DeepLesion \cite{yan2018deeplesion,yan2019holistic}. We have retrained the MULAN models using both CT and EF inputs, but even the best results, {\it{i.e.}}, using EF, have a large gap ($72.5\%$ vs. $78.2\%$ mRecall) with our distance-gating networks, which further proves that the tumor distance transformation cue plays a key role in GTV$_{LN}$ identification. 

\begin{figure}[t]
\centering
\includegraphics[width=0.9\textwidth]{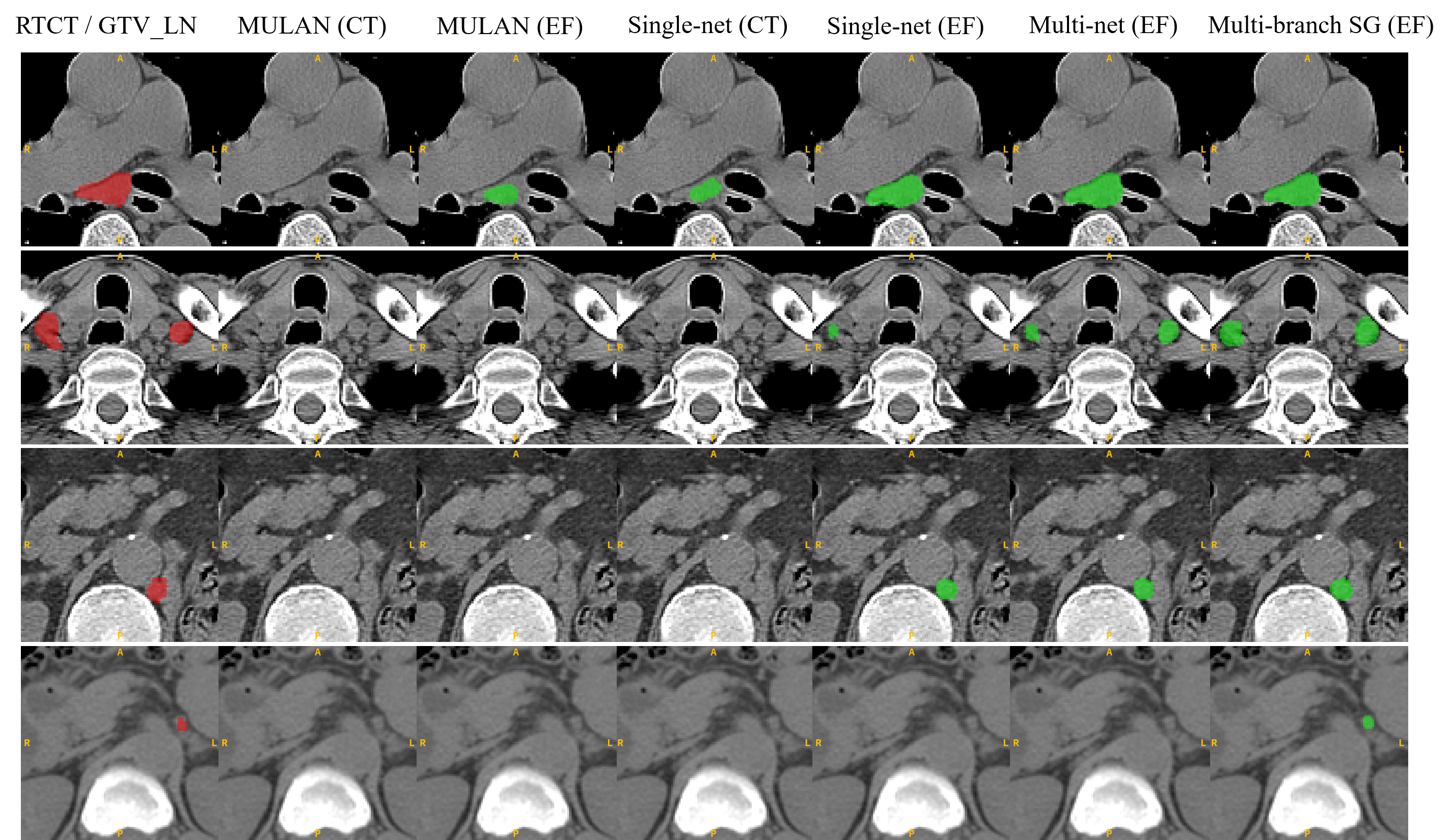} 
\caption{Four qualitative examples of the detection results using different methods. Red color represents the ground-truth GTV$_{LN}$ overlaid on the RTCT images; Green color indicates the predicted segmentation masks. As shown, for the enlarged GTV$_{LN}$ (top row), most methods can detect it correctly. However, as GTV$_{LN}$ size becomes smaller and contrast is poor, our method can successfully detect them while others struggled. } \label{Fig:Examples}
\end{figure}

Fig.~\ref{Fig:Examples} illustrates the visualization results of our method compared to other baselines. For the enlarged GTV$_{LN}$ (top row), most methods can detect it correctly. However, as the size of GTV$_{LN}$ becomes smaller and the contrast is poorer, our method can still successfully detect them while others struggled.

\section{Conclusion}
In this work, we propose an effective distance-based gating approach in a multi-task deep learning framework to segment GTV$_{LN}$, emulating the oncologists' high-level diagnosis protocols. GTV$_{LN}$ is divided into two subgroups of ``tumor-proximal" and ``tumor-distal", by means of binary or soft distance gating. A novel multi-branch detection-by-segmentation network is trained with each branch specializing on learning one subgroup features. We evaluate our method on a dataset of $141$ esophageal cancer patients. Our results demonstrate significant performance improvements on the mean recall from $72.5\%$ to $78.2\%$, as compared to previous state-of-the-art work. The highest achieved GTV$_{LN}$ recall of $82.5\%$ at the $20\%$ precision level is clinically relevant and valuable.

\bibliographystyle{splncs04}
\bibliography{typeinst}
\end{document}